\documentclass[5p,times,twocolumn]{elsarticle}
\usepackage{amsmath,amssymb,amsfonts}
\usepackage{algorithmic}
\usepackage{graphicx}
\usepackage{textcomp}
\usepackage[dvipsnames]{xcolor}

\usepackage{miller}
\usepackage{amsmath}
\usepackage{bm}
\usepackage{siunitx}

\def\BibTeX{{\rm B\kern-.05em{\sc i\kern-.025em b}\kern-.08em
    T\kern-.1667em\lower.7ex\hbox{E}\kern-.125emX}}

\usepackage{hyperref}
\usepackage{cleveref}

\newcommand{\revision}[1]{#1}

\newcommand{\tnsr}[1]{\ensuremath{\mathbf{#1}}}
\newcommand{\F}[1][]{\ensuremath{\tnsr{F}^{#1}}}
\newcommand{\Fp}[1][]{\ensuremath{\tnsr{F}_\text{p}}}
\newcommand{\Fi}[1][]{\ensuremath{\tnsr{F}_\text{i}}}
\newcommand{\Fe}[1][]{\ensuremath{\tnsr{F}_\text{e}}}
\newcommand{\Lp}{\ensuremath{\tnsr{L}_\text{p}}}
\newcommand{\Li}{\ensuremath{\tnsr{L}_\text{i}}}
\newcommand{\shear}[1]{\ensuremath{\gamma^{#1}}}
\newcommand{\A}[1][]{\ensuremath{\tnsr{A}^{#1}}}

\makeatletter
\def\ps@pprintTitle{%
  \let\@oddhead\@empty
  \let\@evenhead\@empty
  \let\@oddfoot\@empty
  \let\@evenfoot\@oddfoot
}
\makeatother

\begin{document}

\begin{frontmatter}

\title{Subgrain-resolved Analysis of Degradation in Cu Metallization via Scanning 3DXRD and Thermomechanical Modeling}

\author[MTM,CS]{N.\,Prabhu\corref{cor1}}
\ead{nikhil.prabhu@kuleuven.be}                                                                       
\cortext[cor1]{Corresponding author}
\author[IKTS]{L.\,Neumann}
\author[KAI]{M.\,Reisinger}
\author[ESRF]{J.\,Ball}
\author[ESRF]{C.\,Corley-Wiciak}
\author[KAI]{M.\,Petersmann}
\author[ESRF]{A.\,Corley-Wiciak}
\author[ESRF]{J.\,Wright}
\author[ESRF]{C.\,Detlefs}
\author[MTM,CS]{M.\,Diehl}

\address[MTM]{Department of Materials Engineering, KU Leuven, Leuven, Belgium}
\address[CS]{Department of Computer Science, KU Leuven, Leuven, Belgium}
\address[IKTS]{Fraunhofer - Institut für Keramische Technologien und Systeme, Dresden, Germany}
\address[KAI]{KAI – Kompetenzzentrum Automobil- und Industrieelektronik GmbH, Villach, Austria}
\address[ESRF]{European Synchrotron Radiation Facility, Grenoble, France}

\begin{abstract}
Metallization layers play a key role in the performance and reliability of modern power semiconductor devices.
During short-circuit events, rapid heating of power metallization layers induces thermomechanical incompatibility stresses, which may contribute to material degradation and impact device performance.
In this work, potential degradation hotspots associated with thermomechanical loading in Cu power metallization are investigated using a combined experimental--computational approach.
Scanning three-dimensional X-ray diffraction measurements are coupled with thermomechanical crystal plasticity simulations to probe the evolution of grain-resolved plastic deformation during rapid cyclic loading.
This integrated approach provides insight into the microstructural processes governing \revision{degradation hotspot formation, laying the groundwork for future microstructure-informed, physics-based reliability assessment of Cu metallization.}
\end{abstract}


\end{frontmatter}

\tableofcontents

\section{\textbf{Introduction}}
The continued downscaling of power semiconductor devices has led to increased current densities.
In the event of a short circuit, dissipation of electrical power results in rapid heating of the device, inducing significant thermomechanical loading in metallization layers.
As these conductive layers act as pathways for electrical current and contribute to heat dissipation, their integrity is essential for device performance and reliability.
Consequently, degradation of metallization layers is considered one of the dominant failure mechanisms limiting the lifetime of power semiconductor devices.

A major source of mechanical loading in power metallization layers that are in direct contact with the semiconductor arises from the mismatch in their thermal expansion behavior \cite{LutzEtAl2018}.
In particular, the thermal expansion coefficient of copper is typically approximately five to six times larger than that of common semiconductor materials such as Si, SiC, or GaN \cite{LutzEtAl2018}.
In extreme cases such as short-circuit events, rapid heating induces significant thermomechanical incompatibility stresses within the metallization layer.
These stresses can exceed the elastic limit of the metallization material, leading to localized plastic deformation.
Furthermore, the associated thermal transients can be ultrafast, with heating rates approaching \qty{1e6}{\kelvin\per\second} \cite{NelhiebelEtAl2013}.
Repeated exposure to such extreme loading conditions can drive progressive localization of plastic deformation and the formation of microstructural defects within the metallization.

While the impact of such degradation processes on device performance is well recognized, the underlying mechanisms remain difficult to access experimentally at relevant spatial scales, as metallization layers are embedded structures containing thousands of grains with diameters of a few micrometers.
Previous studies based on surface-sensitive techniques such as electron backscatter diffraction (EBSD) have been used to track microstructural evolution during thermal cycling; however, with increasing cycling, surface roughening leads to deteriorating indexing quality, limiting reliable characterization \cite{Bigl2017}.
These studies indicate that plastic localization and defect formation evolve at the grain and subgrain level \cite{ZiegelwangerEtAl2025,HlushkoEtAl2023}, requiring subgrain-resolved information across large grain populations within bulk material, which remains inaccessible to conventional surface-based techniques.

Consequently, a synergistic combination of advanced non-destructive characterization and microscale multiphysics modeling---both capable of capturing grain- and subgrain-scale features---is essential to obtain complementary insight into the mechanisms governing thermomechanical degradation within metallization layers relevant for high-power electronics. In this work, we investigate potential thermomechanical degradation hotspots in copper contacts under ultrafast loading representative of short-circuit events.
To this end, scanning three-dimensional X-ray diffraction (s3DXRD) \cite{HenningssonEtAl2024} measurements at the materials science beamline ID11/ESRF \cite{WrightEtAl2020} are carried out at different stages through the device lifetime, and combined with thermomechanical crystal plasticity simulations.
\revision{The novelty of the present work lies in the integrated experimental--computational approach, which enables the correlation of grain-resolved plastic deformation with degradation hotspot evolution during rapid thermal cycling, providing new mechanistic insight into grain-scale processes governing Cu metallization degradation in power semiconductor devices.}

\section{\textbf{Non-Destructive Bulk Characterization}}
\label{sec:Non-Destructive Bulk Characterization}
\begin{figure}
    \centering
    \includegraphics[width=\linewidth]{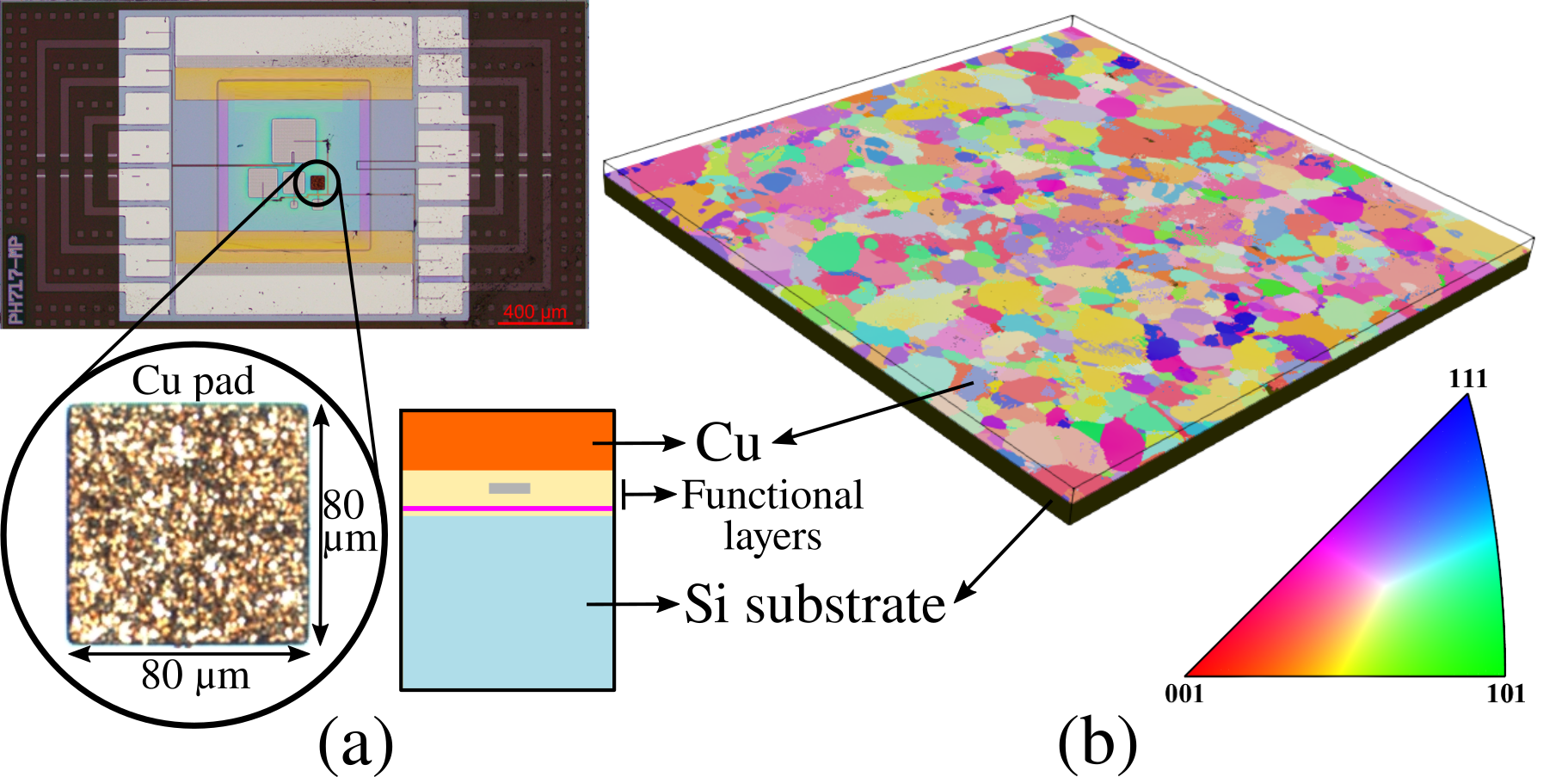}
    \caption{(a) Dedicated test chip with the Cu pad highlighted, including a zoomed-in view and schematic cross-section of the layer stack. (b) Simulation geometry of the Cu metallization on the Si substrate, with grain structure visualized using IPF-Z coloring along the surface normal}
    \label{fig:sample}
\end{figure}

A dedicated test chip (see \cref{fig:sample}~(a)) was previously designed to investigate metallization reliability under service-relevant conditions \cite{MoserEtAl2019}.
The sample consists of an \qtyproduct{80 x 80 x 10}{\micro\meter} electrochemically deposited Cu pad on a \qty{120}{\micro\meter} Si substrate, including intermediate functional layers for Joule heating and temperature sensing.
Post-deposition, the Cu pad was annealed at \qty{400}{\degreeCelsius} for \qty{30}{\min} to stabilize the microstructure.

s3DXRD experiments were performed using a focused X-ray nanobeam (\qtyproduct{150 x 300}{\nano\meter}, \qty{43}{\kilo\volt}).
The measurement strategy combined raster scanning with tomographic rotation at each scan position, enabling non-destructive mapping of the microstructure.
\revision{The raw diffraction data thus acquired at each scan position were segmented, indexed, and refined using processing routines implemented in the ImageD11\footnote{\url{https://github.com/FABLE-3DXRD/ImageD11}} Python library and described in \cite{HenningssonEtAl2024}.
Reconstruction was performed using a point-by-point indexing approach, in which each voxel was reconstructed independently based on diffraction peaks geometrically consistent with its position, without imposing neighborhood-based constraints during indexing.}

A lateral step size of \qty{150}{\nano\meter} was used, corresponding to a \revision{reconstructed} voxel size of \qtyproduct{150 x x 150 x 300}{\nano\meter}.
\revision{Following orientation refinement, the reconstructed microstructures were post-processed using the Python interface of simplnx\footnote{\url{https://github.com/BlueQuartzSoftware/simplnx}} to remove spatially isolated orientation outliers arising from the voxel-wise reconstruction procedure.}
A cross-sectional layer near the mid-height of the Cu pad was selected as the region of interest, and the corresponding inverse pole figure map along z-direction (IPF-Z) is shown in \cref{fig:sample}~(b).

The non-destructive nature of s3DXRD allowed the same sample to be characterized at multiple stages (225, 550, and 1000 cycles) during thermomechanical cycling, in addition to the pristine state.
\emph{Ex situ} thermal loading was applied via Joule heating, with temperature cycles between \qtyrange{100}{400}{\degreeCelsius}, a pulse duration of \qty{100}{\micro\second}, and a dwell time of \qty{1}{\second}.
\revision{The loading conditions were selected to reproduce the rapid thermomechanical loading associated with repetitive short-circuit events, following the polyheater methodology described in \cite{MoserEtAl2019}.}

\section{\textbf{Grain-scale Thermomechanical Modeling}}
Thermomechanical crystal plasticity simulations were performed on the experimentally characterized 2D central layer to complement the microstructural observations from the experiment.
The simulations were carried out using the spectral solver implementation in DAMASK Multiphysics \cite{RotersEtAl2019}, with thermal and mechanical, i.e. elastoplastic, fields solved in a staggered, self-consistent manner.

\revision{The material behavior is modeled in a finite strain framework in which the total deformation gradient $\F$ is multiplicatively decomposed into elastic, inelastic, and plastic contributions (subscripts e, i, p, respectively):
\begin{equation}
  \F = \Fe \Fi \Fp.
\end{equation}
The rate of the plastic deformation gradient can be calculated from the plastic velocity gradient $\Lp$ as
\begin{equation}
  \dot{\Fp} = \Lp \Fp,
\end{equation}
where the crystal plasticity formulation restricts contributions to $\Lp$ to shears $\shear{\alpha}$ on $\alpha = 1,\ldots,N_\mathrm{slip}$ slip systems defined by their Schmid matrix $\tnsr{P}^\alpha$ 
\begin{equation}
	\Lp = \sum \limits_{\alpha=1}^{N_\mathrm{slip}} \tnsr{P}^\alpha \shear{\alpha}.
\end{equation}
Similarly, the inelastic deformation due to thermal expansion is modeled as
\begin{equation}
	\Li = \dot{T} \A,
\end{equation}
where $\dot{T}$ is the temperature rate and $\A$ is the anisotropic thermal expansion tensor.
The stress---arising from thermal expansion in the present context---that drives plastic deformation in the Cu pad is calculated using the generalized Hooke's law for anisotropic materials.
It is complemented by a phenomenological crystal plasticity description \cite{RotersEtAl2019,PrabhuEtAl2024} in which the evolution of the critical resolved shear stress per slip system and the rate-dependent shear are governed by a power-law relation.}

The geometric model, shown in \cref{fig:sample}~(b), consists of a Cu pad modeled as a contact metallization layer on a purely elastic silicon substrate.
To address the periodic boundary conditions inherent to the spectral solver, an air-like material layer \cite{MaitiEtAl2018a} was introduced around the Cu pad to effectively break periodicity, thereby better approximating the physical configuration.

\revision{The temperature history measured using the polyheater setup (see \cref{sec:Non-Destructive Bulk Characterization}) was directly prescribed as the thermal boundary condition for both the Cu metallization and the Si substrate.
Consequently, the simulations primarily investigate the grain-resolved thermomechanical response arising from the mismatch in thermal expansion.
This choice of loading condition is consistent with the available experimental observations, which are limited to a single reconstructed layer near the mid-height of the Cu pad.
}

\revision{The raw experimental data, processed datasets, and the complete computational workflow are publicly available to facilitate reproducibility \cite{CorleyWiciakEtAl2028}.
}

\section{\textbf{Results and Discussion}}
\revision{The experimental and simulation results are presented and discussed in this section to elucidate the microstructural evolution under thermal cycling.}

\begin{figure*}
    \centering
    \includegraphics[width=\textwidth]{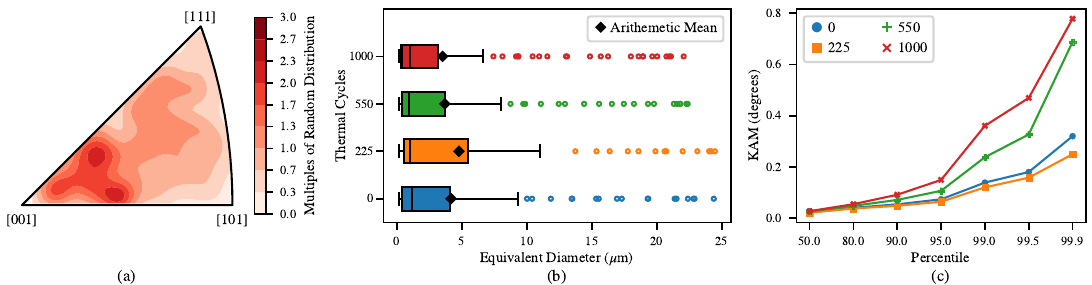}
    \caption{(a) IPF-Z density plot of the pristine Cu metallization layer. Evolution of equivalent grain diameter distributions (b) and selected KAM percentiles (c) for the pristine and thermally cycled states}
    \label{fig:IPF_GZ_KAM}
\end{figure*}
\revision{The crystallographic orientation distribution of the characterized central layer in the pristine state is shown as an IPF density plot in \cref{fig:IPF_GZ_KAM}~(a).
The plot represents the distribution of crystallographic directions aligned with the sample normal (Z-direction), reduced into the fundamental zone of the face-centered cubic crystal symmetry and projected onto a (flattened) sphere.
Although the Cu pad was subjected to a post-deposition annealing treatment, the reconstructed orientations suggest the presence of a weak \hkl<001> fiber component, indicated by the increased density of orientations near the \hkl[001] corner of the standard triangle.
The same can also be visually inferred from the corresponding IPF-Z map shown in \cref{fig:sample}~(b) where orientations associated with the vicinity of the [001] direction are prevalent throughout the reconstructed area.
As the analysis is based on a single reconstructed layer containing a limited number of grains, this observation should be interpreted with appropriate caution.
}

\begin{figure*}[!t]
    \centering
    \includegraphics[width=\textwidth]{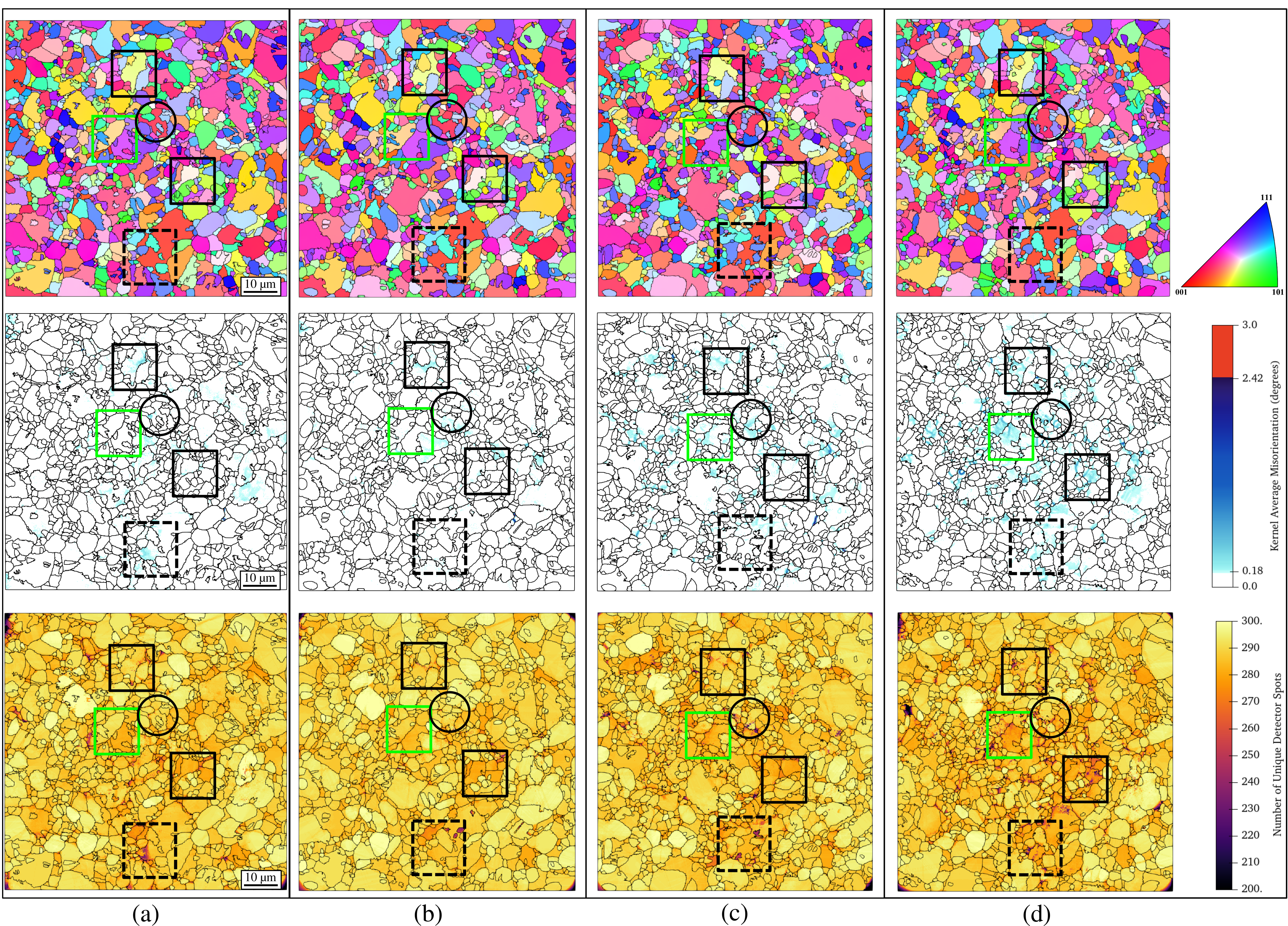}
    \caption{Experimental IPF-Z (top), KAM (middle), and unique detector spots per voxel (bottom) maps for the pristine sample (a) and after thermal cycling to 225 (b), 550 (c), and 1000 (d) cycles. Highlighted areas indicate regions of interest for further analysis}
    \label{fig:MS_maps}
\end{figure*}

\revision{As previously described, the thermomechanical loading was performed \emph{ex situ} between successive s3DXRD characterizations.
Consequently, the exact same physical layer could not be guaranteed for all measurements.
Therefore during each characterization step, the scan layer was selected as close as possible to the mid-plane of the \qty{10}{\micro \meter} thick Cu pad.
In order to assess the comparability of the analyzed layers, the corresponding IPF-Z maps are juxtaposed in \cref{fig:MS_maps}~(a--d).
A qualitative inspection reveals that many characteristic grain morphologies appear consistently across all thermal cycling stages.
Although minor differences arising from repositioning cannot be excluded, the overall similarity of the maps indicates that the analyzed sections originate from comparable regions of the microstructure.
This provides confidence that the trends presented hereafter primarily reflect microstructural evolution rather than being dominated by artifacts due to sample misalignment.}

\revision{A comparison of the grain size distributions, quantified in terms of equivalent grain diameter obtained from the orientation maps, is shown through the box plots in \cref{fig:IPF_GZ_KAM}~(b).
For all investigated states, the median grain size remains close to \qty{1}{\micro \meter}, whereas the arithmetic mean is several times larger.
This difference indicates a strongly right-skewed distribution consisting of a large population of small grains together with a limited number of substantially larger grains.
Comparison across the thermal cycling stages reveals an overall reduction in the higher side of the grain size distribution, reflected primarily by a decrease in the arithmetic mean while the median remains comparatively stable.
A deviation from this general trend is observed after 225 cycles, where the distribution exhibits a temporary increase before decreasing again at higher cycle numbers.}

\revision{We note that previous studies have shown that thermomechanical cycling can induce low-angle boundary formation and hierarchical subdivision of Cu grains through progressive lattice reorientation \cite{ZiegelwangerEtAl2025}.
As these orientation gradients develop, regions that initially belong to the same grain may exceed the segmentation tolerance (chosen here to be \qty{5}{\degree}) and therefore be reconstructed as separate grains as grain identification is based on orientation segmentation in post-processing.
The apparent reduction in equivalent grain diameter may therefore reflect increasing intragranular subdivision in addition to other physical microstructural changes \cite{HlushkoEtAl2023}.
Nevertheless, to minimize any processing-related bias, identical reconstruction and post-processing parameters were applied to all thermal cycling stages.
Moreover, the observed trend in equivalent grain diameter remained unchanged when the grain segmentation misorientation tolerance was varied between \qty{3}{\degree} and \qty{15}{\degree}, indicating that the observed trends are unlikely to be artifacts of the chosen analysis parameters.}

\revision{To gain further insight into the degradation accompanying thermomechanical loading, a complementary statistical analysis based directly on the reconstructed orientation field was therefore performed.
Although s3DXRD does not directly resolve dislocation densities or substructural features, kernel average misorientation (KAM) is commonly used as an indicator of intragranular lattice distortion associated with plastic deformation.
To this end, KAM was computed from the reconstructed orientation field by averaging the local lattice misorientation up to second-neighbor voxel relationships.}

\revision{Because deformation is expected to localize within only a limited fraction of the microstructure, the KAM distribution is anticipated to be concentrated at low values with a long tail extending towards higher misorientations.
To capture this behavior, selected percentiles of the KAM distribution are plotted for each thermal cycling stage in \cref{fig:IPF_GZ_KAM}~(c).
A clear heterogeneous evolution of KAM is observed during thermal cycling.
Between the pristine and the 1000 cycles states, the median KAM increases only modestly from approximately \qty{0.02}{\degree} to \qty{0.03}{\degree} , whereas the 99.9\textsuperscript{th} percentile increases over two times from \qty{0.32}{\degree} to \qty{0.78}{\degree}.
The substantially stronger increase at the higher end of the distribution indicates that thermomechanical loading has a non uniform impact on the state of the microstructure \cite{HlushkoEtAl2023}.
Consequently, the accumulation of orientation gradients becomes increasingly concentrated within a limited fraction of the analyzed section, indicating the presence of highly localized hotspots driving overall thermal degradation.}

\revision{Interestingly, the KAM evolution is not monotonic.
After 225 thermal cycles, KAM decreases across essentially the entire distribution before increasing again at higher cycle numbers.
The same trend is observed when voxels close to the reconstructed sample boundaries are excluded from the analysis, suggesting that it is unlikely to arise solely from reconstruction artifacts at the sample boundaries.
A similar turning point was previously observed in the equivalent grain diameter statistics, suggesting that both descriptors may have captured a common microstructural transition occurring during the early stages of thermal cycling.}

\revision{While the percentile analysis reveals the statistical evolution of KAM, the scanning-probe nature of s3DXRD enables direct visualization of where these orientation gradients develop within the microstructure.
To complement this information, maps of the number of unique diffraction spots contributing to the reconstruction of each voxel are shown in \cref{fig:MS_maps}~(a--d) as a qualitative measure of diffracted signal strength.
Such information is included because thermomechanical fatigue may promote void formation and increasing lattice distortion which reduce diffraction signal quality.
Since low or missing signal may originate from several causes, this metric is used only to qualitatively explore possible correlations between KAM-derived plastic activity and signal degradation.
}

\revision{Examination of the KAM maps reveals that the evolution of intragranular orientation gradients associated with plastic deformation is spatially heterogeneous.
Rather than exhibiting a uniform increase throughout the analyzed layer, regions of elevated KAM evolve differently during thermal cycling.
While some areas displaying elevated KAM in the pristine condition become less pronounced after 225 cycles, other regions emerge and progressively intensify at higher cycle numbers.
This behavior is consistent with the non-monotonic evolution observed in the statistical analysis.
}

\revision{Although the statistical confidence associated with a single reconstructed layer remains limited, the elevated KAM values observed in the pristine condition may reflect residual orientation gradients inherited from the electrodeposition process.
Similar persistence of low-angle boundaries and intragranular defect structures after post-deposition annealing has been reported for electrodeposited Cu metallization \cite{ZiegelwangerEtAl2025}.
The reduction in KAM after 225 cycles suggests that thermal cycling initially promotes microstructural rearrangement before deformation-induced lattice distortion becomes dominant.
A similar interplay between energy-minimizing microstructural processes and subsequent deformation-driven subdivision has been reported in thermomechanically cycled Cu metallization \cite{BiglEtAl2016,HlushkoEtAl2023,ZiegelwangerEtAl2025}.}

\revision{Representative examples of the trends discussed so far are highlighted in \cref{fig:MS_maps}.
The two solid black boxes identify regions where noticeable KAM is already present in the pristine condition and progressively develops into pronounced hotspots with continued cycling.
In contrast, the green box exhibits only limited activity during the early stages before a rapid increase in KAM after 550 cycles.
The dashed box displays elevated KAM in the pristine state, a reduction after 225 cycles, and a subsequent re-emergence at higher cycle numbers, mirroring the non-monotonic evolution observed in the statistical analysis.}

\revision{The diffraction spot maps reveal a concurrent reduction in signal strength within and in the vicinity of some of the marked regions where a significant change in KAM was observed.
Most notably, the area marked by the black circle exhibits a pronounced signal decrease after 550 and 1000 cycles.
Although reduced signal strength alone cannot confirm void formation, its spatial proximity to regions of elevated KAM is noteworthy.
Under the same thermal cycling conditions generated using the polyheater methodology, void nucleation and growth have been reported within the first 1000 thermal cycles, predominantly at grain boundaries and triple junctions \cite{KleinbichlerEtAl2021}.
The present observations may therefore indicate the early stages of local damage evolution, consistent with previous studies linking localized deformation substructures to vacancy accumulation and subsequent microvoid nucleation in Cu metallization \cite{NoellEtAl2020,HlushkoEtAl2023}.}

\revision{Taken together, the experimental observations indicate that thermomechanical cycling progressively concentrates deformation-induced lattice distortion into a limited number of microstructural regions.
The concurrent reduction in detector signal observed near some of these hotspots suggests that they may also become preferential sites for damage initiation, although confirmation of the underlying mechanism requires further investigation.}

\begin{figure*}
    \centering
    \includegraphics[width=\textwidth]{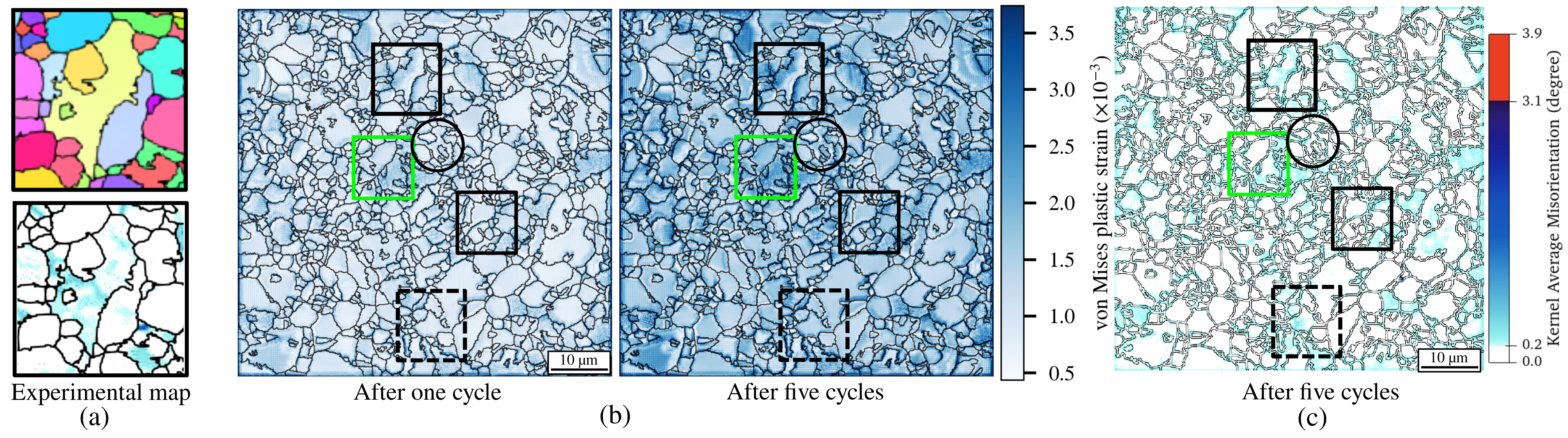}
    \caption{(a) Zoomed-in images of the region highlighted by the black box on the top in \cref{fig:MS_maps}~(d), corresponding to 1000 cycles. (b) Equivalent von Mises plastic strain and (c) KAM maps obtained from cyclic thermomechanical simulations using the experimental microstructure; boxes correspond to regions in \cref{fig:MS_maps}.}
    \label{fig:sim_MS_maps}
\end{figure*}
To further interpret the experimentally observed trends, plastic deformation obtained from the thermomechanical simulations is analyzed.
While KAM serves as a proxy for plastic activity in the experiments, the simulations enable direct quantification through the plastic component of the strain field, expressed here as the von Mises equivalent plastic strain.
The corresponding maps of von Mises plastic strain and KAM across two different thermal cycling stages are shown in \cref{fig:sim_MS_maps}~(b, c).

A strong spatial correspondence is observed between the regions highlighted in the experimental maps (\cref{fig:MS_maps,fig:sim_MS_maps}~(a)) and those exhibiting elevated plastic strain and KAM in the simulations.
Both measures indicate progressive localization of plastic activity with thermal cycling, with existing hotspots intensifying and additional localized regions emerging at higher cycle counts.
The simulations additionally predict elevated plastic deformation at grain boundaries compared to grain interiors, consistent with previous experimental observations \cite{HlushkoEtAl2023}.
\revision{Furthermore, plastic strain localization is not distributed uniformly among all grain boundaries, suggesting that hotspot formation is influenced not only by the properties of an individual grain but also by its local neighborhood and grain boundary configuration.}

\revision{The observed qualitative agreement between experiment and simulation largely supports the interpretation that the experimentally detected hotspots in lattice distortion are associated with localized plastic deformation.
It should be noted, however, that the employed constitutive description is phenomenological in nature and describes plastic deformation through dislocation-mediated crystallographic slip.
Furthermore, the constitutive description does not explicitly account for recovery processes, dislocation substructure formation, vacancy evolution, or damage formation.
The model is therefore used here to investigate the mechanisms governing plastic strain localization and degradation hotspot formation rather than to establish a quantitative correspondence between experiment and simulation.
The chosen model will always predict a monotonic accumulation of plastic strain with increasing thermal cycling and can therefore not reproduce the apparent turning points observed experimentally.
Consequently, the simulated cycle number should not be interpreted as directly corresponding to the experimental number of thermal cycles.
Similarly, the experimentally observed intragranular band-like features, indicative of low-angle boundary formation, are not explicitly resolved within the current constitutive formulation.}

\revision{Nevertheless, the good qualitative agreement between simulations and experiments demonstrates that crystal plasticity-based microstructure models provide a reliable framework for identifying regions susceptible to plastic strain localization and for guiding microstructure-informed design and reliability studies of Cu power metallization.
Such a microstructure-sensitive approach could ultimately contribute toward physics-based lifetime assessment of Cu metallization subjected to repetitive short-circuit loading.
To that end, capturing such subgrain formation may further require deeper physics-based descriptions of the underlying defect structure and evolution.
Furthermore, complementary information on void locations and local stress states in the vicinity of these hotspots---the latter accessible through s3DXRD together with the three-dimensional microstructure---represents a promising direction toward improved mechanistic understanding of degradation in Cu metallization.}

\section{\textbf{Conclusion and Outlook}}
Repeated short-circuit events in power semiconductor devices induce thermomechanical stresses in Cu power metallization, leading to a progressive increase in intragranular orientation gradients, indicative of the accumulation and localization of plastic deformation.
This evolution is characterized by both the intensification of existing hotspots and the emergence of new regions of elevated plastic activity with increasing thermal cycling.
\revision{In addition, a non-monotonic evolution is observed during the early stages of cycling, where both KAM and grain size statistics exhibit a turning point, suggesting that microstructural recovery or rearrangement processes may initially compete with deformation-induced substructure formation.}

A concurrent reduction in detector signal intensity is observed in regions exhibiting elevated KAM.
While this correlation suggests a possible link between localized deformation structures and the onset of damage evolution, reduced signal can also arise from data processing limitations and therefore cannot be unconditionally attributed to void formation based on the present data alone.
Complementary characterization, such as nanoCT or 3D EBSD on thermally cycled samples, is therefore required to reliably identify voids in order to elucidate their formation mechanisms.

\revision{Overall, the combined analysis of orientation gradients, detector signal intensity, and plastic strain localization provides complementary insight into the microstructural processes governing degradation in Cu metallization and establishes a foundation for identification of potential degradation hotspots.
However, the experimentally observed non-monotonic evolution is not reproduced by the simulations, highlighting the importance of recovery and non-local defect evolution mechanisms that are not accounted for in the employed constitutive description.
Although plastic strain localization appears to be closely associated with the onset of damage evolution, a complete description of the underlying degradation process additionally requires consideration of the local stress state.
Future work will therefore combine the elastic strain information accessible from s3DXRD with complementary void characterization and enhanced constitutive descriptions incorporating additional physical mechanisms to establish a more complete link between stress evolution, plastic strain localization, and damage formation in Cu metallization.}
Advancing this understanding remains an important step toward microstructurally informed, physics-based reliability assessment frameworks for power electronic devices.

\section*{Acknowledgment}
We acknowledge the European Synchrotron Radiation Facility for provision of synchrotron radiation facilities under proposal number MA-6791 and we would like to thank the staff for assistance in using beamline ID11.
This study has received financial support from the European Union: AddMorePower (GA 101091621).
Views and opinions expressed are however those of the author(s) only and do not necessarily reflect those of the European Union.
Neither the European Union nor the granting authority can be held responsible for them.
The computational resources used in this work were provided by the VSC (Flemish Supercomputer Center), funded by the Research Foundation - Flanders (FWO) and the Flemish Government.

{\footnotesize \bibliography{references}}

\end{document}